\documentclass[
reprint, amsmath, amssymb, pre
]{revtex4-2}
\usepackage{graphicx}
\usepackage{dcolumn}
\usepackage{bm}
\usepackage{tikz}
\usepackage{hyperref}
\usepackage{xcolor}
\usepackage{mdframed}
\usepackage{natbib}
\DeclareRobustCommand\dashed{\tikz[baseline=-0.6ex]\draw[thick,dashed] (0,0)--(0.54,0);}
\begin{document}

\title{Confinement-driven state transition and bistability in schooling fish}% Force line breaks with \\

\author{Baptiste Lafoux}
\author{Paul Bernard}
\author{Benjamin Thiria}
\author{Ramiro Godoy-Diana}
\email{ramiro@pmmh.espci.fr, bthiria@pmmh.espci.fr}
\affiliation{
Laboratoire de Physique et Mécanique des Milieux Hétérogènes (PMMH), CNRS UMR 7636, ESPCI Paris—PSL Research University, Sorbonne Université, Université Paris Cité, 10 rue Vauquelin, 75005, Paris, France
}

\begin{abstract}
We investigate the impact of confinement density (i.e the number of individuals in a group per unit area of available space) on transitions from polarized to milling state, using groups of rummy-nose tetra fish (\textit{Hemigrammus rhodostomus}) under controlled experimental conditions. We demonstrate a continuous state transition controlled by confinement density in a group of live animals. During this transition, the school exhibits a bistable state, wherein both polarization and milling states coexist, with the group randomly alternating between them. A simple two-state Markov process describes the observed transition remarkably well. The confinement density influences the statistics of this bistability, shaping the distribution of transition times between states. Our findings suggest that confinement plays a crucial role in state transitions for moving animal groups. More generally, they provide an experimental benchmark for active matter models of macroscopic, self-propelled, confined agents.
\end{abstract}

\maketitle

%#########################################################################
% Introduction
%#########################################################################

\section{Introduction}

Active systems are a broad family of physical systems in which individuals can self propel, through some form of energy consumption. In many cases, collective states can emerge from the interactions between individuals. Within this framework, out-of-equilibrium physics can study a variety of biological systems with a large number of agents, and introduce phenomenological models that can predict and interpret the collective behavior of complex living systems. Conversely, biological systems often serve as experimental systems for physicists looking to study collective dynamics, providing a testing ground for various physical models. This interplay has been fruitful ever since the founding works of Vicsek \cite{vicsek1995}. Collective behaviors of macroscopic biological systems are diverse but, studies but can be classified into three distinct families \cite{calovi2014a}: swarming, a disorganised state with isotropic individual orientations; polarization, in which members of the group are aligned with each other; and milling, a state where individuals move forming a vortex-like structure.

The transition to this milling state is widespread and has been extensively described for a number of different species: marine worms \cite{franks2016}, planktonic crustaceans \cite{ordemann2003}, army ants \cite{schneirla1944}, reindeers \cite{espmark2002}, and most notably in fish \cite{harvey-clark1999, couzin2002, wilson2004, lukeman2009}. In some groups of animals, this milling structure is part of a multi-stable regime, where the system switches between several states, depending on the experimental conditions \cite{strombom2022, d.strombom2022}. Collective motion confers a variety of advantages to the group, from improved predator avoidance and escape \cite{inada2002, ioannou2012, ioannou2017} to a reduced cost of locomotion \cite{hemelrijk2015, ashraf2017, li2019}.

Collective states can be affected by various physical parameters, and some are able to trigger state transitions in groups of moving organisms by altering the nature or intensity of the inter-individual interactions, such as light intensity \cite{ordemann2003, lafoux2023, xue2023} or noise \cite{jhawar2020}. The existing numerical models of self-propelled particles (SPP) also indicate that one of the key factors explaining these transitions is the density of the group \cite{ordemann2003, vicsek1995, biancalani2014, dyson2015}. A common way to modify the density is through confinement, whether by boundaries or a confining potential. Properties of active systems tend to be influenced by the presence of boundaries, such as their dynamic behavior \cite{ribeiro_active_2016} or spatial distribution \cite{yang_aggregation_2014, fily_dynamics_2014, kudrolli_swarming_2008, liu2020,galajda_wall_2007}. In other systems, simulated or experimental, confinement can lead to an altered phase space, with particle segregation \cite{yang_aggregation_2014}, emergence of collective motion \cite{caprini_collective_2021, kudrolli_swarming_2008, wioland2016, beppu2017, mehes2014}, suppression of  phase separation \cite{ben_dor_disordered_2022}, or formation of new phases \cite{rana_tuning_2019}. From these different cases it is clear that, as in other areas of physics, the role of confinement can be central to the behaviour of an active system, even in the thermodynamic limit \cite{ben_dor_disordered_2022}. However, there appears to be no study, theoretical or otherwise, of the effect of confinement on macroscopic biological systems.

For fish schools specifically, studies have highlighted density-driven transitions, either experimentally \cite{becco2006, tunstrom2013} or in simulations \cite{cambui2012, cambui2018}.  These studies primarily investigate the role of the school size, while also noting that the proximity to walls can play a crucial role in triggering specific transitions, like the milling to polarization transition \cite{tunstrom2013}. We can unify these observations by examining the issue of state transition from the perspective of group confinement ---in the sense of how crowded the available area is--- which includes two distinct factors: the number of individuals in the group, and the surface area of the swimming arena.

The study by \cite{tunstrom2013} indicates that the swimming area is not significant in the state transitions between swarming, milling, and polarized states in groups of golden shiners. However, this conclusion was based on a single experiment in which the swimming area was changed. As of yet, no study has systematically investigated how confinement, both in terms of the number of individuals and the arena surface, impacts groups of animals. Notably, the link between confinement and behavioral state transitions remains unsettled.

Here we report quantitative results on the collective dynamics of groups of rummy-nose tetra fish (\textit{Hemigrammus rhodostomus}), a highly cohesive fresh water fish, under controlled experimental conditions. In first approximation, if one neglects the exact shape of the tank, the notion of confinement can be simply quantified by a "confinement density" $\rho$, which is the number of fish per unit area of the tank. Our analysis provides novel experimental evidence for a continuous state transition governed by confinement density. During this transition, the school experiences a bistable state, in which both polarization and milling states coexist; the group can fall randomly into one of these two states. Bouts of variable durations of either state intercede one another alternately over time. By measuring the distribution of transition times between states, we show that the statistics of this bistability is also directly influenced by the confinement density. Through these results, we show that the presence of walls and the group size play a comparable role in the emergence of collective dynamics in schooling fish. 

\section{Experimental setup and data processing}

\begin{figure*}
\includegraphics[width=0.8\linewidth]{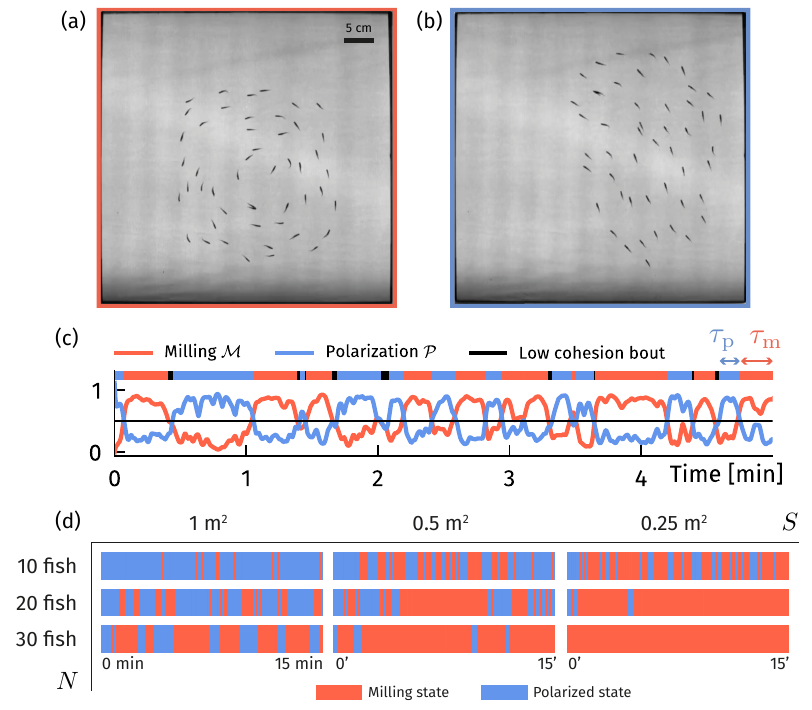}
\caption{Example of an experimental result for $N=50$ fish schooling in a square tank of swimming area $S=1$ m$^2$. Snapshot of the school in (a) milling configuration ($\mathcal M \approx 1$) and in (b) polarized configuration ($\mathcal P \approx 1$). The photographs show the entire surface of the tank. (c) The time signal of $\mathcal M$ and $\mathcal P$ shows clearly distinct periods of milling state followed by periods of polarized state of respective duration $\tau_\mathrm m$ and $\tau_\mathrm p$. The colored bar represents the current state of the school. (d) State transitions over 15 min experiments for various school sizes $N$ and swimming area $S$. [See Supplementary Movie S1.]}
\label{fig:large_figure_photo_timeseries}
\end{figure*}

The experiment consists of recording the motion of free-swimming schools of \textit{Hemigrammus rhodostomus} in a tank, using the same setup as described in \cite{lafoux2023}, for multiple swimming areas $S$ and school sizes $N$ (number of individuals in the group) (see Appendix I for details). We systematically investigated the role of the confinement density, or mean density, $\rho$, defined as the ratio of these two values $\rho = N/S$. The tank is sufficiently shallow (6 cm) to constrain the trajectories in two dimensions. The average body length (BL) of the fish used here is 32 mm (see Appendix I for details on the fish breeding and on the experimental procedure). The swimming area $S$ can be modified by a system of movable partition walls. Three different surfaces are considered: 1, 0.5 and 0.25 m$^2$. We focus mainly on arenas with a square shape (aspect ratio $AR$ = 1), but we briefly discuss the implications of a different aspect ratios. We studied 10 different group sizes $N$ ranging from 10 to 70 fish, resulting in densities between 10 and 120 fish/m$^2$. During the experiments, the tank is brightly lit from above with visible light (900 lux) and back lit with an infrared LED panel to enhance the contrast of the images captured by the camera (Fig.~\ref{fig:large_figure_photo_timeseries}a-b). The camera captures images at 50~frames per second (fps), corresponding to an inter-frame interval of $\Delta t = 0.02$ s.

Using the open-source particle tracking library Trackpy \cite{crocker1996, allan2021}, we extract the temporal signals of the two-dimensional positions $\mathbf{x}_i(t)$, from which we obtain velocities with a second order central differences method $\mathbf{v}_i(t) = [\mathbf{x}_i(t+\Delta t) - \mathbf{x}_i(t-\Delta t)] / 2\Delta t$, where $i=1, 2,..., N$ is the fish label (See Appendix I for details on the tracking accuracy). For each confinement density, at least 2 different pairs of values of $N$ and $S$ corresponding to that density were tested (except for the two extreme density values 10 and 120~fish/m$^2$). We conducted at least 3 different trials of 15 min for every pair of values (swimming area, school size). In total, 76 distinct experiments have been conducted.

We use the canonical milling and polarization parameters ($\mathcal M$ and $\mathcal{P}$) to capture the collective dynamics at the group level:
\begin{equation}
\label{eq:defM}
    \mathcal M = \frac 1 N \left|\sum_{i=1}^N \tilde{\mathbf r}_i \times \tilde{\mathbf v} _i\right|
\end{equation}
\begin{equation}
\label{eq:defP}
    \mathcal P = \frac 1 N \left|\sum_{i=1}^N \tilde{\mathbf v} _i\right|
\end{equation}
where $r_i$ is the position of the $i$-th fish with respect to the school's center of mass ($\mathbf{r}_i = \mathbf{x}_i - 1/N \sum_{i=1}^N \mathbf{x}_i$), and $\tilde{\cdot}$ is the normalization operator ($\tilde{\mathbf u } = \mathbf u  / |\mathbf u|$). These parameters allow the structure of the school to be described mathematically, highlighting the typical states discussed above. When $\mathcal M$ is large (close to 1) and $\mathcal P$ is small (close to 0), the group is in the milling state; conversely, when $\mathcal M$ is small and $\mathcal P$ is large, the fish are in the polarized state. Fig.~\ref{fig:large_figure_photo_timeseries}a-b provides examples of these characteristic schooling states in our experimental setup. \\

%#########################################################################
% Results
%#########################################################################
\section{Results}

\subsection{Bistability of the fish school}

At the smallest density values (10-20 fish/m$^2$), we observe that the school remains in a polarized state for almost the entire recording duration, with only short-lived (a few seconds) incursions to the milling state that quickly revert to the polarized state. This behavior is illustrated in Fig.~\ref{fig:large_figure_photo_timeseries}d, for $S = 1$ m$^2$ and $N=10$. As we further increase the confinement density by adding more fish to the school or reducing the swimming area (typically for densities between 20 and 80~fish/m$^2$), the bistability can be more clearly observed, as milling bouts last longer. The group state shows an increase in fluctuations, with frequent shifts observed between the polarized and milling states, although the proportion of time in the polarized state remains more significant for densities lower than 40~fish/m$^2$; as the density increases again, the milling state becomes more predominant (see Fig.~\ref{fig:large_figure_photo_timeseries}d, for $S = 0.5$ m$^2$ for example). Finally, when approaching densities of 100 fish/m$^2$ and above, the groups mostly display a milling behaviour that lasts for long durations of time, still punctuated with occasional very short periods of polarization (typically a few seconds). This is shown by the panel corresponding to $S = 0.25$ m$^2$ and $N=30$ in Fig.~\ref{fig:large_figure_photo_timeseries}d.

This bistability of the school has been observed previously in experimental \cite{tunstrom2013} and numerical studies \cite{strombom2022, d.strombom2022}, and is reminiscent of a simple dynamical system oscillating between two stable states. This leads us to define these two states objectively: we say that the school is in the milling state (resp. in the polarized state) when $\mathcal M > 0.5$ (resp. when $\mathcal P > 0.5$). The colored bar above the time signal in Fig.~\ref{fig:large_figure_photo_timeseries}c shows that the two states follow each other almost without interruption. {Brief bouts of disordered state ($\mathcal{M}$ and $\mathcal{P}$ both low, on Fig.~\ref{fig:large_figure_photo_timeseries}c in black) are also observed. They happen when the group is in transition or situations where the school is temporarily stuck in a corner of the arena. In the latter analysis, due to the very sporadic nature of these brief periods, we choose to only consider milling and polarized bouts.}

\begin{figure}
\centering
\includegraphics[width=0.83\linewidth]{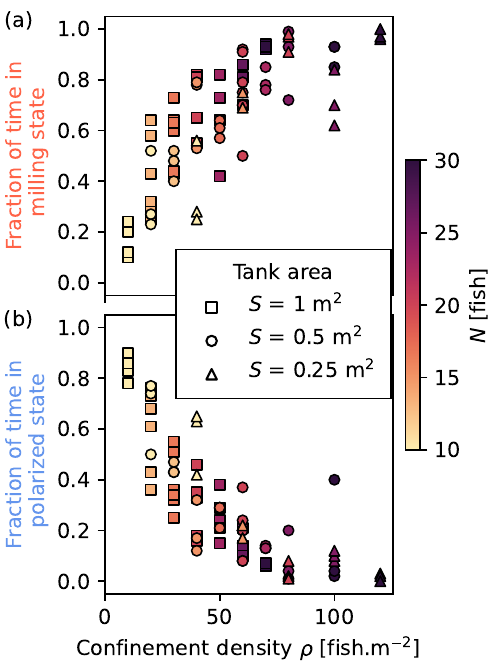}
\caption{Fraction of time spend by fish schools in the milling state, ($\mathcal M>0.5$) (a) and in the polarized state, ($\mathcal P > 0.5$) (b) over total experiment duration, with respect to confinement density, i.e the number of fish per unit area of the tank in fish/m$^{2}$. The size of the markers indicates the area of the tank, while the colors show number of fish in the school.}
\label{fig:master_curve}
\end{figure}

\subsection{Confinement density}

In order to quantify the observed intermittency in states we first investigate it on the time scale of each experiment (15 minutes). Fig.~\ref{fig:master_curve} shows the time spent in either state plotted as a function of confinement density $\rho$. First, it corroborates the observations made on individual experiments, where time spent milling goes up with the density, and oppositely for time spent in the polarized state. Most strikingly Fig.~\ref{fig:master_curve} shows a clear collapse of the data of a total of 76 experiments with 20 different ($N,S$) combinations when plotted against the confinement density. For each confinement density (with the exception of 10 and 120 fish/m$^{2}$), experiments with at least two ($N,S$) combinations were carried out, to ensure that they yielded equivalent results if the $N/S$ ratio was the same. This collapse clearly highlights that the proportion of time spent by the group of fish in either state is solely a function of the confinement density, rather than of the number of fish alone, as suggested by previous studies on golden shiners \cite{tunstrom2013} (see also Appendix II).

\begin{figure}[t]
\includegraphics[width = \linewidth]{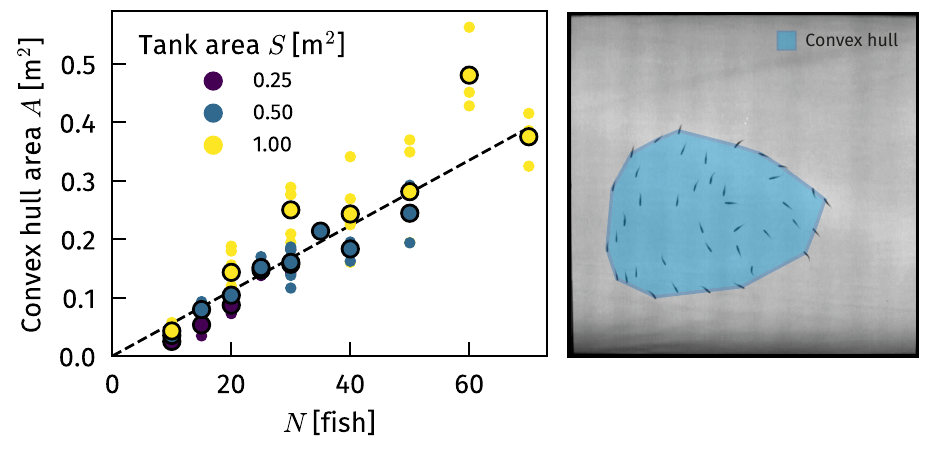}
\caption{\label{fig:grp_size}Area of the convex hull of the school $A$, averaged over the duration of the experiments, expressed in m$^2$, as a function of the numbber of fish $N$. The colors represent the different swimming areas $S$ and the dashed line (\dashed) is the best linear fit for all the data. Small dots represent unique experiments and large points are average over the replicates.}
\end{figure}

In order to assess the local density of the group, we use tracking data to evaluate the temporal evolution of the surface area occupied by the school. To do this, the area of the convex hull of the school $A$ is determined at each time step. This is the area of the smallest convex surface containing all the points $\mathbf{x}_i$ for $i \in 1\dots N$, i.e. the positions of the fish in 2D (an illustration is given in Fig.~\ref{fig:grp_size}). The variations of $A$ (averaged over the duration of the experiments) are shown in Fig.~\ref{fig:grp_size} as a function of the experimental parameters.

%From the trajectories $\mathbf x_i(t)$ obtained by tracking, we extract two geometrical values to gain insight on the organization of the group, the group size $D$ and the Nearest-Neighbor distance NND. Let us denote $d_{ij}$ the distance between the two fish labelled $i$ and $j$ respectively, with $(i, j) \in \{1..N\}^2$. The group size $D$ is obtained by sorting the $N(N-1)/2$ unique distances $d_{ij}$ and taking the average of the 8 largest. If the group is forming a disk (i.e in the milling state), $D$ is a proxy of the diameter of this disk. In the polarized state, $D$ can be understood as the major axis length of an ellipse surrounding the group.

Fig.~\ref{fig:grp_size} shows the values of $A$ with respect to the number of fish in the school $N$, for the 3 different swimming areas $S$. The first observation is that variations of $S$ have no influence on the school area, and that the school area increases slowly with $N$. We fit the experimental data with a linear function 
\begin{equation}
	A = S_0N,
\end{equation}
with $S_0$ a constant fitting parameter homogeneous to a surface. The evolution of $A$ is well captured by this fit, which suggests that whatever the experimental conditions, we can consider that the fish school maintains a constant local density. The best fitting parameter value is found to be $S_0 = 3.67 \pm 1.13 \, \text{ BL}^2$: this surface can be interpreted as a minimal comfort area maintained by each individual around itself. This can also be understood in the following way: the local density is conserved, and is on average 179 fish/m$^2$. 

\begin{figure*}
    \centering
    \includegraphics[width=0.95\linewidth]{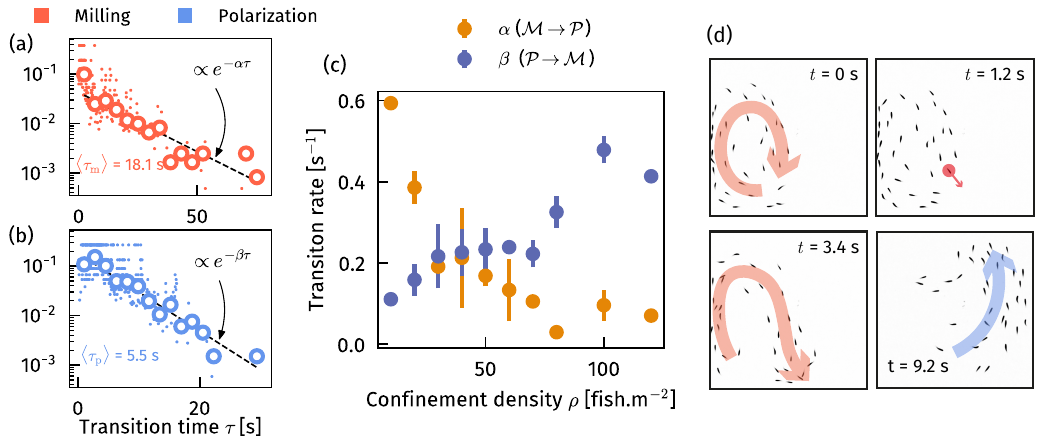}
    \caption{\label{fig:large_figure_model}Probability density function of the duration of milling state $\tau_\mathrm m$ (a) and polarized state $\tau_\mathrm p$ (b) for a confinement density $\rho = 30$ fish/m$^2$ ($N=30$, $S=1$ m$^2$, 6 replicates, 387 transition events). Small dots are generated by differentiating the complementary cumulative density function obtained from sorting the transition times, they do not depend on the binning. Large points are histograms obtained by binning data into 15 linearly spaced time intervals. Dashed lines show the best exponential law fit with parameter $\alpha = \langle \tau_\mathrm m \rangle ^{-1}$ and $\beta = \langle \tau_\mathrm p\rangle^{-1}$ (c) Transition rates with respect to confinement density. Error bars represent the standard deviation. (d) Experimental snapshots of 40 fish in a 1m$^2$ tank illustrating the mechanisms underlying a transition from milling to polarized state. The black borders represent the tank walls.}
\end{figure*}

\subsection{A two-state model}

We introduce a simple two-state Markov process at the scale of the school to gain insight on the transition between the polarized and milling states. In this scenario, the school is considered as a bistable system; we denote $\alpha$ (resp. $\beta$) the rate of transition from the polarized to milling states (resp. from milling to polarized). The transition probability from polarized to milling (resp. milling to polarized) during a time $\delta t$ is therefore $\delta t  \alpha$ (resp. $\delta t  \beta$).  This probability is considered to depend only on the current state and is independent of the history of the system. In this case, the time between switching events (i.e the duration of the bout spent in either state) follows a Poisson distribution with a rate parameter specific to that state. We have the following probability density functions for the duration of milling and polarized bouts:
\begin{equation}
    \mu_\mathrm p(\tau_\mathrm p = \tau) = \alpha \exp(-\alpha\tau)
\end{equation}
\begin{equation}
    \mu_\mathrm m(\tau_\mathrm m  = \tau) = \beta \exp(-\beta\tau)
\end{equation}
We experimentally measured the time spent in the milling and polarized states. The results, reported in Fig.~\ref{fig:large_figure_model}a-b show an example of the empirical distribution of the durations between state transitions, for a confinement density at which fish spend approximately the same time in the milling and polarized states ($\rho = 30$ fish/m$^2$, for $N=30$ and $S=1$ m$^2$). At low densities, where transitions are frequent and the experimental statistics is therefore good, fitting an exponential distribution to the data shows good agreement. It should be noted that as density rises, and milling bouts become very long, the number of transitions falls sharply, and the number of bouts observed is much lower. For these high densities, the fit quality deteriorates, likely due to the insufficient statistics behind the experimental distribution. From the fit parameters we can directly obtain the mean time to transition for each state, with $\langle\tau_\mathrm m \rangle = \beta^{-1}$ and $\langle\tau_\mathrm p \rangle = \alpha^{-1}$. The good agreement demonstrated by the distributions in Fig.~\ref{fig:large_figure_model}a-b, which closely fit to an exponential distribution is the hallmark of a memory-less process. With this approach we extract $\alpha$ and $\beta$ for every set of experiments, that is for all pairs of ($N$, $S$). The variations of the two rates are presented on Fig.~\ref{fig:large_figure_model}c, with respect to the confinement density. We see that the stability of the milling state increases with density, while that of the polarized state decreases.

When considering this two-state description of the fish group, one can look into the mechanisms that lead the system to switch from one to the other. From the bistability studied in our experiments, we can take away a few qualitative observations. The polarization to milling transition seems to be the consequence of the front of the polarized group seeing its back and, like a snake biting its tail, switching to a `circular chase'. The mechanism of the milling to polarization transition is less obvious, but appears to be caused by the behavior of one individual straying away from the mill and entraining the rest of the group leading to the breaking of the mill. This observation is a qualitative confirmation of previous numerical simulations by Calovi et al. \cite{calovi2014b}. This mechanism is illustrated in Fig.~\ref{fig:large_figure_model}d. 

%#########################################################################
% Discussion 
%#########################################################################

\subsection{Confinement aspect ratio}

In this study, density variations have been carried out at a constant aspect ratio $AR$, for square-shaped tanks ($AR=1$). However, since wall interactions were found to play a dominant role in the observed collective state transitions, a series of additional qualitative experiments at different $AR$ was conducted (see Appendix III). We observe that, starting from a situation where milling is predominant at $AR = 1$, it disappears and is replaced by polarization as $AR$ increases, i.e. for increasingly elongated tanks (see Fig.~\ref{fig:changing_AR}). This behavior points interestingly to the fact that, in addition to confinement density, the shape of the tank plays a non-negligible role in the milling-polarization transition described here. This could explain previous experimental results that seem contradictory \cite{tunstrom2013}, where milling decreases with increasing density, but where the ratio aspect is not preserved.

\section{Conclusions}

The experiments reported here constitute the first quantitative laboratory study of the influence of confinement on the social behavior of live animals. Along with the works previously reported in \cite{lafoux2023}, these results demonstrate that the collective states depend both on interactions between individuals and the environment of the fish group. Fundamentally, the existence of some coupling between the collective state and the surroundings is unsurprising, due to environment related evolutionary pressures. However, the quantitative relationship between confinement density and bistability is noteworthy. The confinement intensity appears to play a pivotal role in both the overall proportion of time spent in either state and the statistical nature of bout durations. Furthermore, the qualitative investigation of different aspect ratios has shown that the boundary conditions can remove the bistability and select either state. Understanding the physical ingredients underpinning collective behavior may be the key in reproducing some of the functions in synthetic systems. This way, functions like phototaxis \cite{ben_zion_morphological_2023}, chemotaxis \cite{theurkauff_dynamic_2012}, and fish-inspired schooling \cite{berlinger_implicit_2021} behaviors have been reproduced in laboratory developed systems. The experimental findings reported here may serve as an empirical reference point for future numerical or theoretical models, allowing for comparative analysis and validation.

While fish confinement in this study uses walls, parallels can be drawn with natural systems. Predation-induced confinement varies, from physical boundaries like humpback whales corralling fish \cite{sharpe1997} to effective boundaries as seen with dusky dolphins herding fish into `prey balls' \cite{vaughn2011}. Although qualitative, these analogies offer insights into geometric confinement and predation pressure, opening avenues for research and understanding complex behaviors in the wild.

%#########################################################################
% Appendix / Supplementary Information 
%#########################################################################

\section*{Appendix I: Details on the experimental setup and data processing}

\subsection{Experimental setup}
\begin{figure}[b]
\centering
\includegraphics[width=\linewidth]{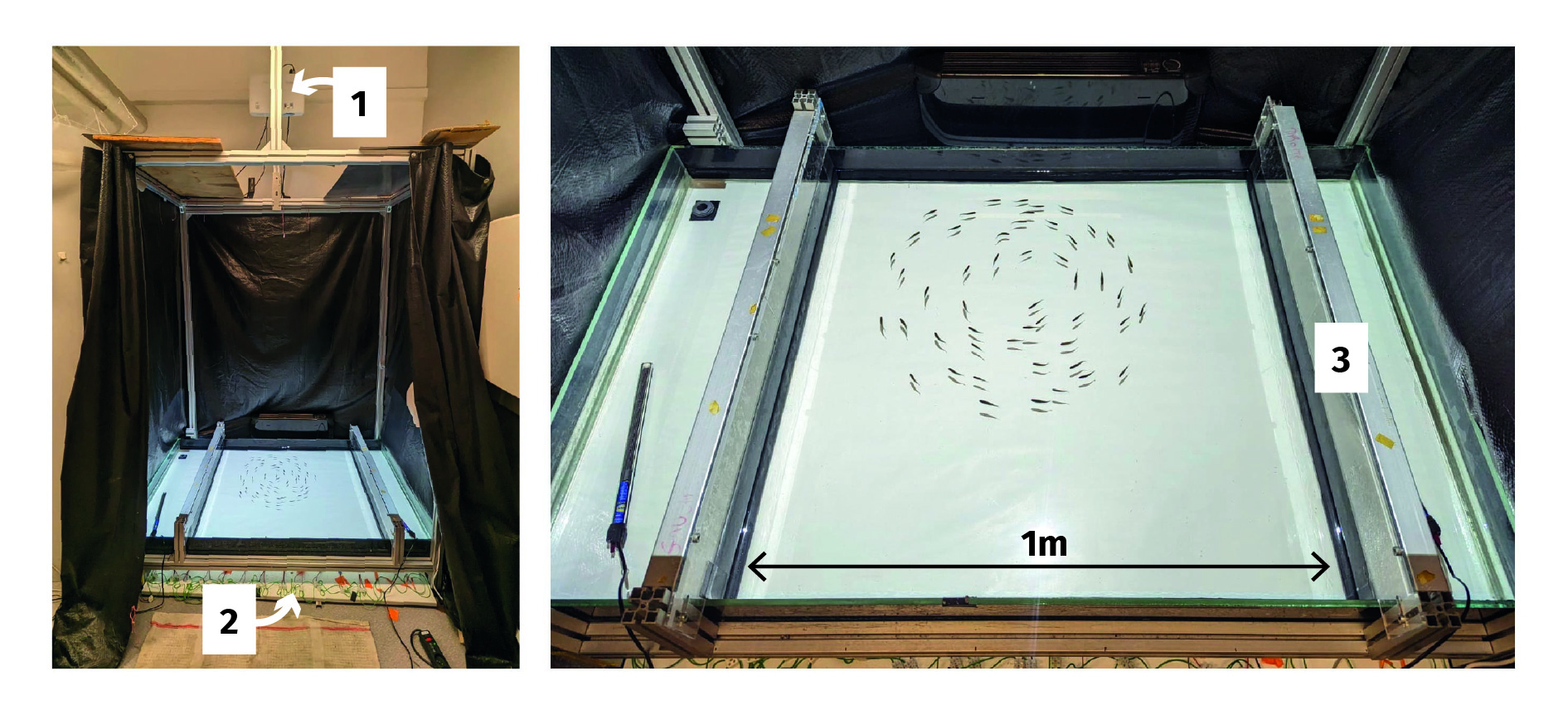}
\caption{\label{fig:exp_setup}Experimental setup. (1) Videoprojector used to set the ambient light intensity. (2) Infrared LED pannel used as a back lighting. (3) Mobile partition walls to set the swimming area.}
\end{figure}
The experimental setup is a shallow water tank (100 cm $ \times$ 140 cm) with mobile partition walls used to modify the swimming area (Fig.~\ref{fig:exp_setup}-3). For areas of 0.5~m$^2$ and 0.25m$^2$, a third partition wall is added to maintain a square tank shape. The resulting side lengths are 68$\pm 2$ cm and 50$\pm$2 cm, respectively. A video projector is mounted approximately 250 cm above, projecting downward to control ambient light intensity with still white images. For all experiments, illumination is set at 900~lux. A custom-made infrared LED panel ($\lambda=890$nm) is placed beneath the tank, enabling high-contrast imaging in any visible light condition without disturbing the fish. A Basler\texttrademark{} camera (4 Mpx, not shown) is positioned 300 cm above the water surface. It's equipped with a visible light filter (passing only infrared light above 920 nm) to eliminate interference from room lighting.

A list of all the $AR$=1 experiments used to produce Fig.~\ref{fig:master_curve} is given in Table~\ref{fig:listofexp}.
\begin{table}[h!]
\centering
\includegraphics[width=\linewidth]{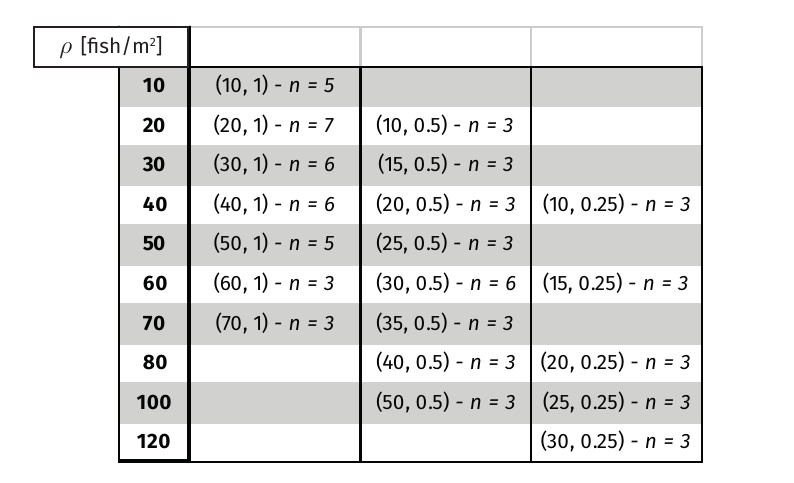}
\caption{List of the experiments carried out for each pair of parameter (number of fish, swimming area [m$^2$]); $n$ is the number of replicates of the same experiment, with a different group of fish.} 
\label{fig:listofexp}
\end{table}

\subsection{Experimental procedure}
\label{ssec:exp_procedure}

\subsubsection{Fish Breeding}
\label{sssec:fish_breeding}
%% PARAGRAPHE COPIE COLLE DE L'ARTICLE LAFOUX ET AL
Rummy-nose tetra fish (\textit{Hemigrammus rhodostomus}) of body length ${\mathrm{BL}=3.9\pm0.4}$ cm were bought from a professional supplier (EFV group, http://www.efvnet.net). Fish were kept in a 120 liter tank on a 14:10 h photoperiod (day:night), similar to that existing at their latitudes of origin. The water temperature was maintained at $27\pm1^{\circ}$C and fish were fed ad libitum with fine pellets from an automated feeder once a day, at a fixed time in the morning. The fish handling protocol complies with the European Directive 2010/63/EU for the protection of animals used for scientific purposes, as certified by the ESPCI Paris Ethics Committee.

\subsubsection{Experiments}
\label{sssec:experiments}

Fish are manually counted and transferred from their aquarium to the observation tank. Manual count of the fish sometimes leads to a small error in fish count, in particular when the total number is high: therefore, for values of $N$ higher than 30, the stated value of $N$ for an experiment should be considered with a $\pm 1$ uncertainty. The aquarium is enclosed by a curtain, so that the projected light intensity (900~lux) is constant from one experiment to the next. Before starting the experiment, the fish are given 10 minutes with no illumination to acclimate to the shallow water tank. The experiment is then carried out and lasts 15 minutes. Experiments can be repeated up to 3 times with 10 minute rest intervals where the projector is switched off. Fish are then given a period of at least 24h to rest between two experiment sessions.

Note that for the smallest values of swimming area, we did not carry out trials with large numbers of fish, as this would have resulted in excessively high densities, which could have been stressful for the animals. On the other hand, we could not reach values of density of density lower than 10 fish/m$^2$ because we have chosen to restrict ourselves to a number of individuals greater than 10, as it is uncertain whether the interactions are similar in a case where very few individuals interact.

\subsection{Fish tracking and data processing}
\label{sssec:fishtracking}
\begin{figure}[b]
\includegraphics[width=0.9\linewidth]{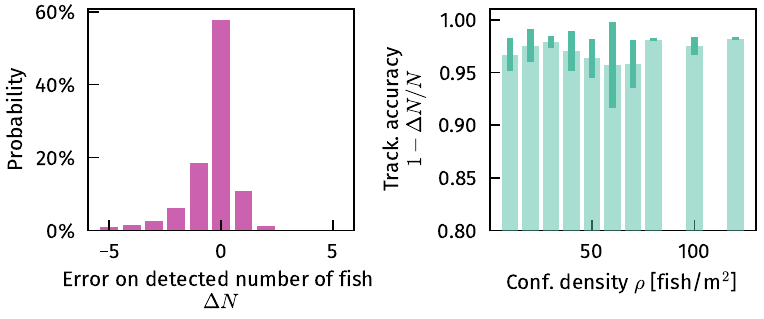}
\caption{\label{fig:track_accuracy} (left) Distribution of the error on the number of fish detected by the tracking algorithm $\Delta N$ (for all the frames in all the experiments). (right) Tracking accuracy (ratio between number of detected fish and fish present in the tank) for the different confinement densities $\rho$ tested in this study. The error bars represent the standard deviation over the different experiments at a given $\rho$.}
\end{figure}

Experiments are recorded at 50 fps and then analyzed using the Python library Trackpy  \cite{crocker1996, allan2021} which returns the positions $\mathbf x_i (t)$ of detected fish at each frame and connects the points on different frames to form trajectories. The position function of each fish is then filtered with a Savitzky-Golay filter of order 2 (window of 21 frames, which corresponds to approximately 0.4 s at 50 fps). The filtered signal is then differentiated to obtain the velocities $\mathbf v_i(t) $ of the fish with a second-order central differences method:
$$
\mathbf{v}_i(t) = \frac{\mathbf x_i(t+\Delta t) - \mathbf{x}_i(t-\Delta t)}{2 \Delta t},
$$
where $\Delta t=0.02$ s is the acquisition period.

The tracking quality is evaluated using two metrics: error in fish detection ($\Delta N$) and tracking accuracy (ratio of detected to actual fish $(N - \Delta N)/N$). Fig.~\ref{fig:track_accuracy} shows the evaluation of the tracking quality. All fish are detected 58\% of the time, with absolute error less than 1 in 88\% of cases. The error distribution ranges from -5 to 5. Tracking accuracy remains above 95\% across all tested confinement densities ($\rho$).

Once the trajectories have been obtained using tracking, the temporal signals of the order parameters $\mathcal M(t)$ and $\mathcal P(t)$ are computed. To assess the behavioural state of the school (milling or polarized), these time signals are filtered using a Gaussian filter with parameter $\sigma$ = 16. In this way, the information contained in the signals is smoothed over a characteristic duration of approximately 1 s. This time scale represents a compromise, which makes it possible to reduce the noise present in the order parameter signals while losing a minimum of information, since the transition times are typically much greater than 1 second. 

\begin{figure}[b]
    \centering
    \includegraphics[width = \linewidth]{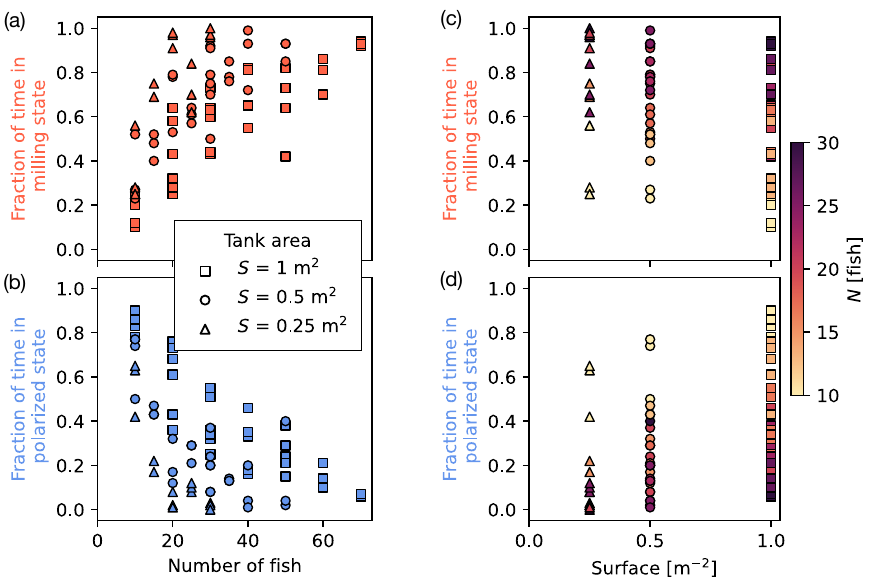}
    \caption{Fraction of time spend by fish schools in the milling state, ($\mathcal M>0.5$) (a, resp. c) and in the polarized state, ($\mathcal P > 0.5$) (b, resp. d) over total experiment duration, with respect to $N$, number of fish in the group, resp. $s$ the surface of the tank. }
    \label{fig:altfig2}
\end{figure}

\section*{Appendix II: Influence of $N$ and $S$ on the time spent in either state}

One of the main findings of the present work is that the time the fish group spends milling or polarized is controlled by the confinement density $\rho = N/S$. To support the claim that $\rho$ is the relevant parameter, we present here the same plots as a function of $N$ or $S$ in Fig.~\ref{fig:altfig2}. In comparison to Fig.~\ref{fig:master_curve}, Fig.~\ref{fig:altfig2} shows that using $N$ or $S$ as the controlling parameter for the time spent in either phase leads to a much larger spread of the data.

\begin{figure*}
\includegraphics[width=0.75\textwidth]{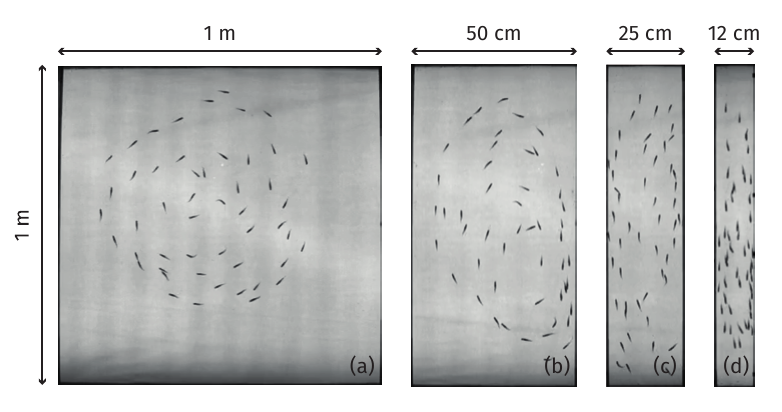}
\caption{\label{fig:changing_AR}Effect of a variation of the aspect ratio $AR$ of the tank, from $AR = 1$ to $AR = 10$. The height of the swimming area remains fixed at 1 m while its width changes: (a) 1 m (b) 0.50 m (c) 0.25 m (d) 0.1 m. Panel (e) shows an experiment in a 0.2 m $\times$ 0.2 m ($AR = 1$). Note that the milling motion is totally suppressed for $AR > 2$.}
\end{figure*}

\section*{Appendix III: Role of the aspect ratio $AR$ of the tank}
To assess the influence of the aspect ratio of the swimming tank $AR$ on the  transition between polarized and milling sates, we conducted additional qualitative experiments with different $AR$ values, specifically $AR$ = \{1, 2, 4, 10\}. In this series of experiments, the length of the tank remains fixed at 1 m, while its width is varied to 50, 25, and 10 cm, respectively. All other experimental parameters, such as frame rate, water depth, and ambient light intensity, were kept consistent with the conditions used in the initial set of experiments. This additional dataset consists of one video recording of 10 minutes of schooling for each value of $AR$. Because of the proximity imposed on the fish in this setup (especially at high $AR$), trajectory crossings are common and periods where fish are superimposed are regularly observed; therefore, tracking accuracy drops drastically and renders quantitative analysis with the tracking pipeline used for this study unreliable. However, a qualitative analysis of collective behaviour is still possible from the raw images. 

As shown in Fig.~\ref{fig:changing_AR}a-d, milling is clearly suppressed for elongated tanks with aspect ratios greater than 2. Starting from a situation where milling exists approximately 60\% of the time in a square tank, we see a reduction in the time spent milling and a deformation of the vortex structure formed by the fish when we move to an aspect ratio of 2. Beyond $AR$ = 2, the milling almost no longer exists (no longer at all for $AR=10$), and the fish move back and forth between the two ends of the tank, remaining aligned along its long side. We note that in this case the transition is exclusively due to geometric constraints, since at the confinement density considered here (respectively 100, 200, 500 fish/m$^2$ for $AR=$ 2, 4, 10), we are well above the threshold after which we observe only milling in the case of a square tank. It is also interesting to point out that, starting from a situation where the milling is suppressed (for example the tank with an aspect ratio of 4), it is still possible to make the milling 'reappear' by reducing the aspect ratio, as demonstrated by an additional experiment carried out in a 20 by 20 cm square tank (see Fig.~\ref{fig:changing_AR}e). 

We hypothesize that these observations explain the experimental results obtained by Tunström et al. \cite{tunstrom2013}: they reported that, when reducing the swimming area, higher confinement density does not lead to increased time spent milling. However, this reduction of area was conducted at an aspect ratio of approximately 2 (66 $\times$ 38 cm), which means that they might have observed the same phenomenon of geometrical suppression of milling that we describe here. 

\acknowledgments{B. L. and P. B. contributed equally to this work.}

%\bibliography{response_bib}

\end{document}